\documentclass{acm_proc_article-sp}

\usepackage{graphicx}  
\usepackage{epsfig}
\usepackage{amssymb}
\usepackage{amsmath}
\usepackage{amsfonts}
\usepackage{algorithm}
\usepackage{algorithmic}

\begin{document}
%
\title{Code Attestation with Compressed Instruction Code}
%
%
\vspace{0.5cm}
\author{Benjamin Vetter, Dirk Westhoff\\	      
		   Fakult\"at Technik und Informatik\\
		   HAW Hamburg \\	
		   Hamburg, Germany \\
		   \{vetter\_b, westhoff\}@informatik.haw-hamburg.de \\
}

%
%

%
\markboth{short paper submission to ACM WiSec'11}{Shell \MakeLowercase{\textit{et al.}}: Bare Demo of IEEEtran.cls for Journals}


\maketitle

\begin{abstract}
Available purely software based code attestation protocols have recently been shown to be cheatable.
In this work we propose to upload {\em compressed} instruction code to make the code attestation protocol robust
against a so called compresssion attack. The described secure code attestation protocol makes use of recently
proposed micro-controller architectures for reading out compressed instruction code.  
We point out that the proposed concept only makes sense if the provided cost/benefit ratio for the aforementioned 
micro-controller is higher than an alternative hardware based solution requiring a tamper-resistant hardware module.
\end{abstract}



\section{Introduction}
The evolution of the ubiquitous computing vision towards full-fledged real world applications faces a diversity of new problems. Besides other issues and due to the fact that for many applications due to the large number of involved end-devices cost-efficient hardware is an issue, one can not guarantee that a code image which once has been uploaded on a tiny, non-tamper resistant device, will always run in a correct and un-manipulated way. Even worse, it may behave in a Byzantine manner such that the device sometimes behaves correctly and sometimes behaves incorrectly. 

One strategy to control respectively detect such misbehaving nodes in a sensor network, or, more generally, in an M2M setting, is to run from time to time a challenge-response protocol between the restricted device and a master device - the verifier - that is sending the challenge.

However, recently it has been shown that purely software based code attestation \cite{SCUBA}, \cite{SWATT}, \cite{ESAS} is vulnerable against a set of attacks. Basically one can subdivide code attestation techniques into two subsets: the first class of approaches is using challenge-response protocols in conjunction with harsh timing restrictions for the restricted device's response. 
Otherwise an attacker could simply load the original code image into the external memory and save program memory
for his own bogus code. Each time the master device triggers the code attestation protocol, for the computation
of the response the cheated prover device reads the original program from the external memory. Since
reading from external memory is much more time consuming, a timing restriction at the verifier for the duration between sending the challenge message
and receiving the response message can detect this.
The second proposed class of countermeasures randomly fills empty program memory to avoid that such free memory space can be used to infect the device with bogus code. 
In their landmark work \cite{CCS09}, Castelluccia et al. have shown that both types of aforementioned countermeasures can be circumvented. Later we provide more details on this.
The rest of the paper is organized as follows: Section 2 introduces the adversary model. Section 3 describes the so called  compression attack the attacker can perform to break recently proposed code attestation protocols. In Section 4 we propose our countermeasure 
to deal with compression attacks and in Section 5 we give insights how to execute compressed instruction code as necessary requirement for this approach. Section 6 discusses suitable compression algorithms and in Section 7 we provide the security analysis of the proposed solution. Conclusions and open issues are presented in Section 8.

\section{Adversary Model}
After node deployment and before the first round of the attestation protocol starts, the attacker
has full control over all device memories such that he can modify program memory or any other memories
like e.g. the external memory. 
At attestation time, when the challenge-response based attestation protocol is running, 
the attacker has no physical control over the restricted device anymore. 
However, please note that the device may yet run malicious code. It is up to the code
 attestation protocol to detect this independently of the fact that the attacker may find ways to store 
the original uploaded code image at a different memory than the program memory.
Note that we do not consider fluctual data memory. \textit{Control Flow Integrity}
could prevent attacks that use techniques like \textit{Return-Oriented Programming} \cite{CCS09}, \cite{controlflowintegrity1}, 
\cite{controlflowintegrity2}.
Obviously, during the phase in which the attacker has full control over the restricted device, 
the attacker is also able to either modify the code for the code-attestation protocol itself or to read out any sensitive data like e.g. pre-shared keys in case the code attestation protocol would be based on this.

\section{Compression Attack}
One major challenge for a purely software-based code attestation for embedded devices is the so called compression attack. This attack cheats a basic challenge-response based code attestation as follows: the originally uploaded program which shall temporarily be checked by the attestation protocol to be exclusively stored in the program memory is subsequently compressed by the attacker. Depending on the concrete compression algorithm and according to the actual uploaded 
code image for a given application the compression gain ranges from 12\% up to 47\% \cite{CCS09}. 
An attacker can use such free program memory to store and run bogus code on the node's program memory. Note that 
current solutions for secure code-attestation also propose to fill the free program memory with pseudorandomly generated words instead of the default entry 0FF. This defends against an attacker who could use
this previously unused memory for uploading a bogus code image in an undetected way.
Since the aforementioned pseudorandomly generated words are required to be part of the response of a code attestation protocol, the verifier needs to know respectively may be able to compute such pseudorandomly generated words.

However, Castelluccia et al. have shown that cheating such kinds of attestation protocols 
is still possible: whenever the restricted device (prover) receives a nonce from the master device 
(verifier) it decompresses the earlier compressed original program on-the-fly
and subsequently computes the hash value $x=h(nonce||CI||$ $PRW)$ by applying the hash 
function $h()$. The $x$ is the checksum respectively the response of the challenge-response
 protocol. The $CI$ denotes the originally uploaded code image and the $PRW$ is the
 pseudo-randomly filled content within the remaining free program memory at load-time. 
Obviously this simple challenge-response based code attestation fails: Whenever 
the prover receives a fresh nonce (the master device initiated the code attestation protocol), 
the attacker {\em decompresses} the compressed $CI$ and writes it into the program memory again.
 This provides all the relevant
 input parameters for the computation of the hash function, namely the $CI$, the nonce, 
and the $PRW$ such that the master device subsequently receives the response $x$ within a given 
time interval which it verifies to be correct.
Finally note that, to save his own bogusly uploaded code image $\widetilde{CI}$, the attacker could have stored $\widetilde{CI}$ also within the external memory.
Subsequently to the time-critical code attestation phase, he has enough time to again compress the $CI$ and read 
 $\widetilde{CI}$ from external memory to program memory. 

\section{Attestation of Compressed\\ Instruction Code}
Our countermeasure against uploading malicious code into the program memory and subsequently not being able to
detect this, re-uses and adapts earlier proposed code attestation protocols \cite{SWATT}, \cite{SCUBA}, \cite{ESAS} by at the same time using 
\begin{itemize}
\item[i.] a hardware extension at the micro-controller, and 
\item[ii.] fulfilling a strict policy for uploading $CI$s into the program memory.
\end{itemize}
This policy is to only upload a yet {\em compressed} code image $C(CI)$ into the 
program memory and to fill the remaining part with $PRW$\footnote{We decided not to compress the $PRW$ since in fact a good choice of the pseudorandomly
filled words can not be compressed anymore. In fact $C(PRW)$ would result in $|C(PRW)| \geq |PRW|$
eventually providing another attack vector to save memory by computing $C^{-1}(C(PRW))$.}. 
Consequently, the attacker cannot allocate such easily free program memory anymore to tracelessly upload malicious code by 
applying the above described compression attack.
Note that with the proposed approach the challenge (a fresh nonce) which goes into the hash computation for every run of the code-attestation anew, enforces the prover to always compute the hash value (response) with a compressed $CI$ and $PRW$ anew.
In our proposed setting the response $x$ thus is computed as
$h(nonce||C(CI)||PRW)$ where the $C$ is a properly chosen lossless data compression algorithm. More details on the properties of the chosen lossless data compression algorithm $C$ and other refinements
on $C(CI)$ will be provided later. The adapted code attestation protocol is shown in 
Figure 1 (Option 1).

Please note that still with our proposed adapted code attestation protocol allowing to upload only 
a compressed code image into the program memory it is essential to enforce a runtime restriction 
as a countermeasure against an attack in which the original code image or parts of it are shifted to the external memory.
We term $\epsilon$ as the duration of the time interval $[t_0,t_1]$ measured by the 
local clock of the verifier.
The $t_0$ denotes the sending time of the challenge $nonce$ and the $t_1$
 denotes the receiving time of the response $x$. 
We emphasize that a proper choice of the threshold $T_{em}$ with $\epsilon < T_{em}$ is 
prover device-dependent to defend the approach against attacks using the \underline{e}xternal \underline{m}emory
of the prover device.
\begin{center}
\begin{figure}
\begin{picture}(200,100)

\put(20,86){\line(0,-1){90}}
\put(100,86){\line(0,-1){90}}

\put(25, 70){\vector(1,0){70}}
\put(95, 10){\vector(-1,0){70}}

\put(113,60){\small{$Option$ $1:$}}
\put(113,50){\small{$x=h(nonce||C(CI)||PRW)$}}
\put(113,30){\small{$Options$ $2a$ $and$ $2b:$}}
\put(113,20){\small{$x=h(nonce||C(CI)||dic||PRW)$}}
\put(113,10){\small{$x=h(nonce||C(CI)||LAT||PRW)$}}

\put(12,40){$\epsilon$}
\put(13,38){\line(0,-1){28}}
\put(13,70){\line(0,-1){23}}
\put(11, 70){\line(1,0){5}}
\put(11, 10){\line(1,0){5}}
\put(2,70){$t_0$}
\put(2,10){$t_1$}

\put(45,78){$nonce$}
\put(58,18){$x$}
\put(0,90){$verifier$}
\put(85,90){$prover$}

\end{picture}
\caption{Derivates of the secure code attestation protocol with lossless data compression algorithm.}
\end{figure}
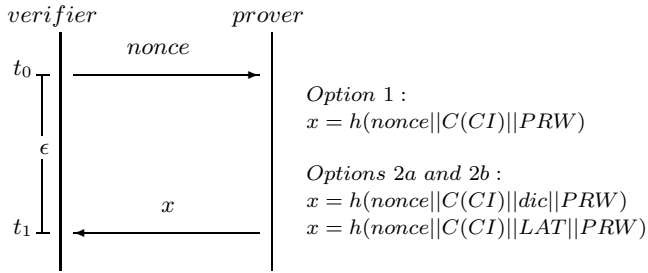
\end{center}

\section{Execution of Compressed\\ Instruction Code}
Now that an attacker cannot such easily cheat the code-attestation protocol anymore by simply compressing the originally uploaded code image
and subsequently decompressing it if needed, the remaining problem with this approach is how to run compressed code? 
To solve this issue one needs to incorporate a hardware extension at the micro-controller. 
Please note that the approach to upload a compressed code image into the program memory is not new. It has recently been proposed by Yamada et al. \cite{patent}. Early work on this can be found in \cite{WoCh}. 

However, originally it has been proposed with the objective to offer a high compression ratio and a
fast instruction expendability - and not 
as a building block to protect against a bogus code image in the program memory like we are proposing. Envisioned is a program memory which 
includes a dictionary memory or other means to start the decompression operation. This component is responsible for storing 
instruction codes which appear in a typical program image. Figure 2 illustrates the micro-controller architecture which
is proposed in \cite{patent}. Another compression technique based on a dictionary has been presented by 
Lefurgy et al. in \cite{LeBi}.
\begin{center}
\begin{figure}
\begin{picture}(200,200)

\put(100,5){\framebox(50,20){}}
\put(100,45){\framebox(50,20){}}
\put(20,45){\framebox(40,20){{\small RAM}}}

\put(140, 130){\line(1,0){10}}
\put(140, 110){\vector(1,0){10}}
\put(130, 93){\line(1,0){20}}
\put(5, 74){\line(1,0){220}}

\put(0,85){\framebox(70,50){}}
\put(5,90){\framebox(60,25){}}

\put(100,85){\framebox(140,100){}}
\put(105,90){\framebox(40,70){}}
\put(150,90){\framebox(80,25){}}
\put(150,120){\framebox(60,40){}}

\put(25,125){\small CPU}
\put(17,105){\small Program}
\put(17,95){\small Counter}

\put(118,56){\small Bus}
\put(107,46){\small Controller}

\put(110,17){\small External}
\put(110,7){\small Memory}

\put(140,175){\small Program Memory}

\put(160,150){\small Compressed}
\put(170,140){\small Code}
\put(165,130){\small Memory}

\put(170,105){\small Dictionary}
\put(168,95){\small (resp. LAT)}

\put(118,130){\small Con-}
\put(115,120){\small troller}

\put(125,44){\line(0,-1){19}}
\put(125,74){\line(0,-1){9}}
\put(37,74){\line(0,-1){9}}
\put(37,84){\line(0,-1){9}}

\put(83,150){\line(0,-1){76}}
\put(83, 150){\vector(1,0){67}}

\put(130,93){\vector(0,-1){19}}
\put(140,130){\line(0,-1){20}}

\end{picture}
\caption{Micro-controller architecture with compressed code memory and dictionary memory \cite{patent}.}
\end{figure}
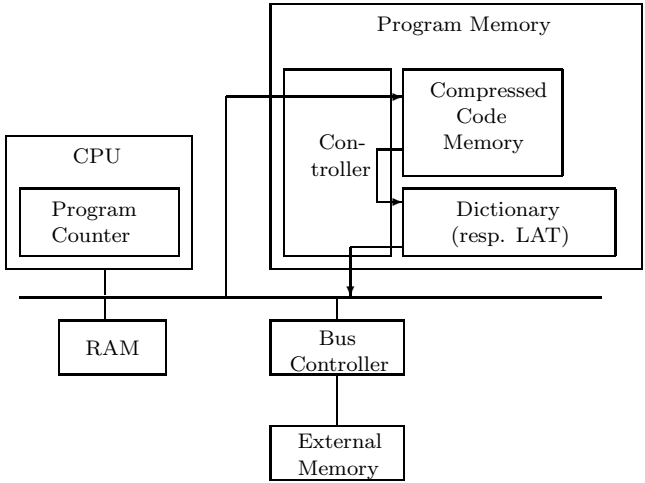
\end{center}
So we propose to only allow to load yet compressed code into the program memory and to decompress the code at runtime. The decompression unit is located at the program memory with a controller passing compressed code instructions to the dictionary memory. This architecture can be used to support the defense against attacks where free program memory space can be generated by compressing the originally uploaded code image and filling this gap with malicious code (including the compression/decompression function).
A code attestation protocol based on simply hashing the original code image plus the remaining free program memory space would not detect such an attack.

Some Remarks: The dictionary memory as well as the compressed code memory are {\em regions} within the
program memory. Thus, in particular the dictionary memory is no dedicated memory module, neither separated nor
protected in a specific manner. Consequently, an attacker could either fully overwrite
or partially modify the dictionary memory. 
To be able to subsequently decompress the $CI$ at runtime we are not allowed to compress
the dictionary ($dic$) itself. We refine the computation of the response $x$ such that:
\begin{equation}
x = h(nonce||C(CI)||dic||PRW)
\end{equation}
This additional consideration of the dictionary has also been reflected within Figure 1 (Option 2a).

\section{Choice of the Data Compression Algorithm}

\subsection{Envisioned Properties}
The proper choice of a suitable lossless data compression algorithm $C$ is essential with respect
to the proposed security architecture. We need to find a lossless data compression algorithm which shall provide the following partially conflicting 
properties:
\begin{enumerate}
\item a high compression ratio for a typical $CI$ (compared to competing lossless data compression algorithms);
\item very fast decompression (vice versa the performance of the compression operation can be relatively poor);
\item the overall decompression concept is required to support entry-points at which 
the decompression operation can start;
\end{enumerate}
With respect to property number one we state that it is one of the properties of any lossless data 
compression algorithm that for {\em typical} input files containing many frequently used 
data chunks the compression rate is rather high. However, vice versa if the input file contains 
many seldomly used data chunks the resulting compression ratio is rather poor. 
Moreover, the compression algorithm $C_h$ chosen by the \underline{h}onest party should ideally provide 
the highest compression rate compared to other compression candidates, e.g. $C_a$ chosen by the \underline{a}ttacker.
Otherwise the attacker could apply $C_a(C_h(CI))$ to save program memory for $\widetilde{CI}$. 

The second property is required since decompression of a code image instruction should
ideally not delay the execution of the originally loaded program. On the contrary there is no technical requirement that restricts the compression time before uploading the $CI$. 

Entry points which define the positions at which the decompression operation starts to decompress the next
code instruction can be either chosen to be placed at fix positions of the compressed $CI$, with a fix and equal distance for a compressed chunk representing a
single code image instruction. This can be achieved by using a dictionary.
A complementary approach would be to allow
entry points at variable positions supporting compression chunks
with different sizes. Clearly the latter provides a better compression ratio
at the cost of a higher management effort for finding the next entry-point.
A cache together with a line address table (LAT) are frequently used for this \cite{WoCh}.
Note that cache and LAT can be independently applied of the concretely chosen compression algorithm.
For this reason we prefer a LAT instead of a dictionary. Our choice has been reflected in Figure 2.

\subsection{Candidates}
Initially we considered {\em Canonical Huffman Encoding} (CHE) \cite{Huf} as 
lossless data compression algorithm $C$ with canonical Huffman tree.
To handle entry points at variable positions with the objective to provide a higher compression rate
we use a LAT as a list of entry points.
Note that with this approach a dictionary memory is not required anymore such that in Figure 1 Option 2b becomes valid:
\begin{equation}
x = h(nonce||C(CI)||LAT||PRW)
\end{equation}
Also, since each entry is listed only one time within the LAT, later we show that 
the attacker does not succeed in sufficiently compressing the LAT. 
It turns out that to a large degree this is also true in case the attacker 
tries to compress the canonical Huffman tree.
However, the disadvantage of the CHE for our purposes is its relatively small gain of compression results on MicaZ with on average $12.19\%$ for various typical WSN programs \cite{CCS09}. For comparison, the lossless data compression algorithm {\em Prediction by Partial Matching} (PPM)  provides an average gain of $47.45\%$ for typical WSN applications. 
Unfortunately, such a significant gain difference of
the compression algorithms $CHE$ and $PPM$ again opens the door for an attack to make use of this gain difference of approximately $35\%$.
The attacker can apply $PPM$ on the compressed code image $C_{CHE}(CI)$ and again generate free space for his own bogus malicious code in either of the two ways: \\
1. $C_a(C_h(CI)) := C_{PPM}(C_{CHE}(CI))$, respectively \\
2. $C_a(C_h^{-1}(C_h(CI))) := C_{PPM}(C^{-1}_{CHE}(C_{CHE}(CI)))$ \\
\noindent
$C^{-1}$ denotes the decompression operation.
Due to the aforementioned reason we also analyzed Deflate, ZPAQ and further derivates of PPM, namely PZIP and PPMZ.
Please note that the hardware supported compression scheme proposed by Wolf et al. \cite{WoCh} doesn't limit the set of lossless compression algorithms.
It only limits the blocksize $s_h$, which has to be equal to the available cache size ($s_h = |cache|$).
\begin{center}
  \begin{figure}[t]
    \includegraphics[width=0.4\textwidth]{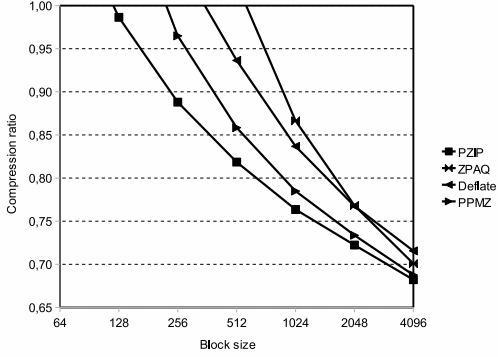}
    \caption{Compression ratios for multi-hop oscilloscope program image of typical compression algorithms for varying blocksizes. }
    \label{compression_ratios}
  \end{figure}
\end{center}
\begin{center}
  \begin{figure}[t]
    \includegraphics[width=0.4\textwidth]{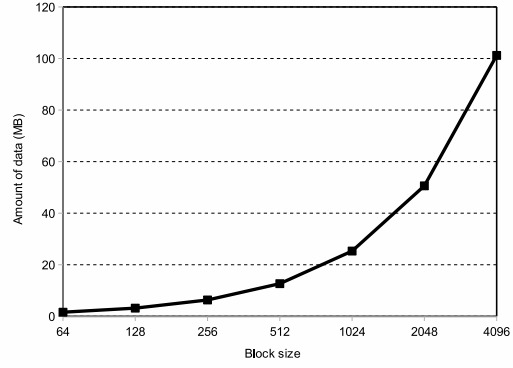}
    \caption{Amount of decompressed data for varying block sizes during the attestation.}
    \label{datavolume}
  \end{figure}
\end{center}
Figure  \ref{compression_ratios} shows that the chosen algorithms provide varying compression
 ratios depending on the block size $s_h$. This is illustrated for our
benchmark code image {\em multi-hop oscilloscope} ($|CI|=25.9KB$) which ships with TinyOS.
Large block sizes provide better compression ratios than small block sizes.
If we choose and apply a tuple $(C_h, s_h)$ the attacker 
can only gain additional free memory $|C_a(C_h(CI))| - |C_h(CI)|$ $ = |\widetilde{CI}|$ by choosing
\begin{enumerate}
\item $s_a>s_h$ if $C_a=C_h$, or 
\item otherwise:  $s_a \leq s_h$ (for some $(C_a, s_a)$).
\end{enumerate}
Nevertheless, if the attacker chooses a much smaller block size the compression ratio will suffer. 
Therefore, when we compress the $CI$ with a larger block size the attacker is forced to use a larger block size as well.
Since the decompression of larger blocks increases the overhead, the time necessary for decompression is increased as well, especially on low-performance platforms like sensor nodes.
This fact becomes significant if we take into account that the attestation has to run
 in a pseudrandomly manner with $nonce$ as the seed for a PRNG 
forcing a strict ordering of the $CI$'s words
when calculating the response $x$ \cite{randomfill}. It forces the attacker to decompress 
each block approximately $s_a$ times.
Moreover, this disables the attacker to apply a 
compression algorithm $C_a$ that sacrifices performance for higher compression ratios since the overhead
increases for larger block sizes $s_a$ recognizably.
Therefore, the use of such algorithms is easily detectable with 
the choice of a large block size $s_h$ and a threshold $T_{pm}$ as the upper
duration for performing compression attacks on the \underline{p}rogram \underline{m}emory. 
Obviously, $\epsilon <  min\{T_{em}, T_{pm}\}$ with $T_{pm} > T_{em}$ as we will see.


Figure \ref{datavolume} shows the amount of temporarily decompressed data during the attestation, 
which increases for larger block sizes.
The attacker has to read about $s_a \cdot |C_{a}(CI)|$ bytes from the program memory during the attestation if he compressed the full $CI$ previously.
If the attacker chooses the block size to be $s_{a}=2048$ bytes and $C_a$ to be PZIP, he will 
have to read up to $37MB$ from program memory to decompress all blocks $s_a$ times and 
subsequently be able to calculate $x$. 
This is a huge overhead compared to $|CI|=25.9KB$. 
For the attacker, obviously this huge amount of data is an immense burden in particular on platforms with low bandwidth for reading from program memory. 
While platforms capable of reading $50MB/s$ result in less than $1$ second timing overhead 
for $2048$ byte blocks, platforms capable of reading only $1MB/s$ require up to $40$ seconds and thus are easily detectable by the proposed attestation protocol.

Obviously, these overhead to decompress every block $s_{a}$ times impacts the time necessary
 for the attacker to calculate the valid response $x$ for the attestation protocol significantly
 on restricted platforms.
As an uncompromised node doesn't have to calculate $C_{h}^{-1}(C_{h}(CI))$ at attestation time, 
i.e. decompress the compressed program image, the block size enables us to raise and adjust the overhead
 for the attacker by orders of magnitude to let us discover the existence of the attacker reliably through a proper choice
 for the device-dependent value of $\epsilon$.
However, a larger cache size respectively $s_{h}$ slow down the on-the-fly decompression routine 
during normal operation of the restricted device.
On the other hand a larger cache decreases the number of cache misses.
Therefore a necessary decompression is more seldom for a larger cache size, but takes more time to complete.

\section{Security Analysis}
Our security analysis considers six attack vectors, namely 7.1 decompressing the code image, 
7.2 attacks on the LAT, 7.3 attacks by using the external memory, 
7.4 replay attacks, 7.5 node depletion attacks, and, finally
7.6 DoS attacks.

\subsection{Decompressing the Code Image}
The attacker is able to decrease the timing overhead by exploiting the fact that different blocks can be compressed with different compression ratios. 
Therefore, the attacker could pick only those blocks which provide the best compression 
ratios out of all blocks until he gains sufficient memory to store his bogus code.
Since the blocks are yet compressed with a properly chosen lossless compression algorithm, each of them provides a similar compression ratio.
To overcome this issue, the attacker could first calculate $C_{h}^{-1}(C_{h}(CI))$, i.e. decompress the compressed $CI$ and compress it for his own afterwards, i.e. calculate $C_{a}(C_{h}^{-1}(C_{h}(CI)))$.
During the attestation he then has to calculate $C_{h}(C_{a}^{-1}(C_{a}(CI)))$ to pass the attestation.
Therefore this method further increases the overhead for the attacker, 
especially if we choose a $(C_h, s_h)$ that compresses rather slowly.
Moreover, the attacker's possible gain is expected to be low, because blocks which provide a good compression ratio to the attacker will provide a good compression ratio to us as well.

\begin{center}
  \begin{figure}[t]
    \includegraphics[width=0.45\textwidth]{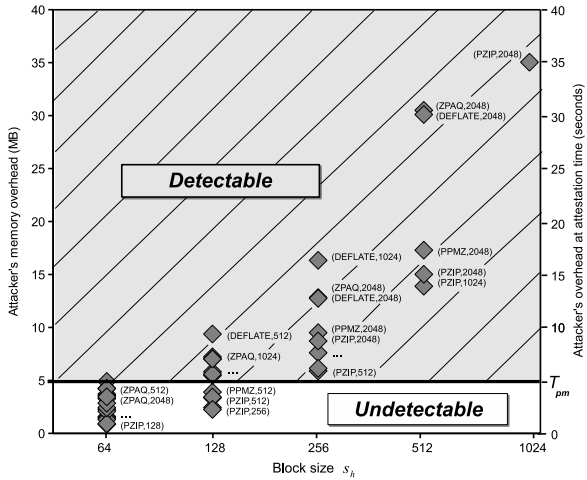}
    \caption{The attacker's possible compression choices for $C_h = PZIP$, a varying $s_h$ and a platform capable of reading $1MB/s$ 
from program memory.}
    \label{pzip}
  \end{figure}
\end{center}

However, even without calculating $C_{h}^{-1}(C_{h}(CI))$ the attacker still requires to compress
 only as much blocks as he needs to gain enough free memory for the $\widetilde{CI}$.
The exact number of blocks an attacker has to use depends on our choice of $(C_h, s_h)$
as well as the attacker's choice $(C_a, s_a)$ and, obviously $|\widetilde{CI}|$ itself.
Please note that besides the $\widetilde{CI}$ the attacker has 
to also store the code of the decompression routine $C^{-1}_{a}$ and 
the $LAT_{a}$ within the program memory. 
As Castelluccia et al. have to spend 1707 bytes for a huffman decompression routine \cite{CCS09}
used in their compression attack, which is a relatively simple algorithm compared to 
the compression algorithms proposed in this paper, we force the attacker to compress
at least multiple blocks to get a chance to gain enough space for his needs.
In general, the attacker has to compress
\begin{equation}
\#Blocks=\frac{|\widetilde{CI}| + |C^{-1}_{a}| + |LAT_{a}|}{GainPerBlock}
\end{equation}
where
\begin{equation}
GainPerBlock=\frac{TotalGain}{\#Blocks_{total}}
\end{equation}
on average with 
\begin{equation}
TotalGain=|C_{h}(CI)|-|C_{a}(CI)|
\end{equation}
and
\begin{equation}
\#Blocks_{total}=\frac{|CI|}{s_{a}}.
\end{equation}
The memory overhead then is about $\#Blocks \cdot s_{a} \cdot \frac{|C_{a}(CI)|}{|CI|} \cdot s_{a}$.
We assume the attacker has to store at least $1KB$ of data\footnote{Please note that 
this is a very optimistic value from the attacker's point of view.}, 
i.e. $|\widetilde{CI}| + |C^{-1}_{a}| + |LAT_{a}| = 1KB$ and
 he will calculate $C^{-1}_{h}(C_{h}(CI))$ before compressing 
$CI$ for his own. 
For example, if we choose $(C_{h}=PZIP, s_h=512$ bytes$)$ 
and the attacker chooses $(C_{a}=PPMZ, s_a=2048$ bytes$)$ 
the attacker's memory overhead is about $17.3MB$.
Figure \ref{pzip} shows the attacker's possible choices for $(C_a, s_a)$ 
to gain sufficient memory whereas $C_h=PZIP$ with varying $s_h$ is our choice of a compression algorithm.
For the attacker's choices we focus on compression algorithms mentioned
 in this paper only, namely PZIP, PPMZ, ZPAQ and Deflate for block sizes ranging 
from $64$ bytes to $2048$ bytes.
On platforms capable of reading $1MB/s$ of data from program memory, we argue
 that memory overhead above $5MB$ is easily detectable since it slows down the attestation
 for about $5$ seconds.
Therefore even if we choose rather small block sizes of $s_{h} \ge 256$ bytes the attack
is still detectable.
Please note that we do not even take the CPU overhead into account here.
From a security point of view we argue to always use the largest possible block size $s_{h}$. 
In practice cache sizes above $1KB$ are hardly feasible, 
especially on embedded devices with less than $4KB$ of data memory.
Therefore we propose to choose $(C_h,s_h \ge 512$ bytes$)$.
Please note that other combinations will be totally feasible as well, 
but one has to choose $s_{h}$ for other compression algorithms more carefully.
\subsection{Attacks on the LAT}
The countermeasure to the compression attack is the compression of the $CI$ with
a suitable data compression algorithm as discussed in Section 7.1. Thus, the only remaining non-compressed data 
besides the $PRW$ which has been argued to be not effectively compressable is the $LAT_h$. 
Consequently, if the $(C_h, s_h)$ for compressing the $CI$ has been chosen properly, 
the only remaining compression attack is to compress the $LAT_h$ itself 
to save program memory $(C_a(LAT_h))$. If the attacker succeeds in saving enough program memory out of this 
to additionally store a bogus code image $\widetilde{CI}$ and at the same time 
requires $\epsilon < min\{T_{em},T_{pm}\}$, the attack is successful and not detectable
 by our code attestation protocol.  
However, recall that a lossless data compression algorithm does not provide the 
same compression ratio for 
every ingoing uncompressed data; in particular a $LAT$ due to its condensed form
can not significantly be compressed as we will see. Moreover, we state that
typically it holds $|LAT_h|<<|CI|$ and $|CI| \leq |PRW|$\footnote{Typical $CI$ sizes for WSN applications are between 10 to 60KBytes such that the $|PRW|$ occupies between 63 Kbytes to 113 Kbytes \cite{CCS09}.}. 
In general, the number of entries of a $LAT$ can be computed as 
\begin{equation}
\#Entries(LAT) = \frac{|CI|}{s}
\end{equation}
So, even if $C_a(LAT_h)$ and $C_a(CI)$ with $(C_a, s_a)$ would provide the same compression ratio, 
which obviously is not the case, the absolute gain of program memory for an attacker who
purely can compress the remaining uncompressed $LAT_h$ would be significantly smaller. 
E.g. we assume an embedded device with $128KB$ of program memory where $|CI|=25.9KB$ (\textit{multi-hop oscilloscope}).
We further assume single $LAT$ entries to be coded using $24$ bits, i.e. $|LAT_h| = \#Entries(LAT_h) \cdot 3$ bytes for the proposed block size $s_h = 512$ bytes.
The $LAT_h$ then occupies $153$ bytes.
Compression results for the $LAT_h$ of our benchmark applications are listed in Table I.
For this setting and by applying our
countermeasure an attacker's absolute gain of free program memory to upload a bogus code image 
$\widetilde{CI}$ would shrink below $5$ bytes approximately\footnote{The attacker can choose other
 compression algorithms not mentioned in this paper as well. Although unlikely, other algorithms could provide slightly better compression ratios.} whereas in the absence of our
proposed solution the attacker could occupy approximately up to $17KB$ of the program memory without being detectable.

Again, with a larger choice of the block size $s_h$ one could reduce
 the free memory space for an attacker even more. 
Furthermore, in case the $CI$ is smaller also the $LAT_h$ shrinks. 
E.g. if $CI$ is  the \textit{BaseStation} respectively \textit{Sense} application and the block 
size again is $s_h=512$ bytes, the attacker will not gain free memory by compressing the $LAT_h$ of size $90$ respectively $18$ bytes using the compression algorithms mentioned in this paper.
Finally, the attacker could overwrite the $LAT_h$ within the program memory for his
 own bogus code; in equivalence to the other program memory containing
compressed code and $PRW$ this attack is detected by the computation and subsequent verification of 
$x=h(nonce||C_h(CI)|| LAT_h|| PRW)$.
\begin{table}[hbp]
\caption{Maximum sizes of bogus code images $|\widetilde{CI}|$ for $s_h=512$ bytes and various applications.}
{\small
\begin{tabular}{|l|c|c|c|}
\hline
         & \textit{Multi-hop os-} & \textit{BaseStation} & \textit{Sense} \\
         & \textit{cilloscope} [byte] & [byte] & [byte]\\
\hline \hline
$|CI|$  & 25906 & 15240 & 2860 \\
\hline 
$|LAT_h|$ & 153 & 90 & 18 \\ 
\hline
$|PZIP(LAT_h)|$ & 148 & 92 & 30 \\
$|PPMZ(LAT_h)|$ & 163 & 109 & 48 \\
$|Deflate(LAT_h)|$ & 181 & 123 & 48 \\
$|ZPAQ(LAT_h)|$ & 242 & 188 & 131 \\
\hline
max. $|\widetilde{CI}|:$ & & & \\
1. our approach & 5 & 0 & 0 \\
2. Refs. \cite{DRSA}, \cite{ESAS} & 16948 & 7029 & 1124 \\ \hline
\end{tabular}}
\end{table} 
\subsection{Attacks using External Memory}
The proposed solution detects attacks by the usage of external memory with the introduction 
of a device-dependent threshold $\epsilon < T_{em}$.
Since the threshold should be as harsh as possible there will definetively be cases in which
a false negative will be the result of a single code attestation run.
Nevertheless we recommend to choose the $T_{em}$ as harsh as possible to indeed have a meaningful 
countermeasure against an attack in which the attacker makes use of the external memory. As a consequence,
in case of a false negative one should repeat the code attestation protocol $n$ times where
$n$ is factor two the number of protocol runs in which the received $x$ does not match to the computation at 
the verifier. 
To restrict the number of iterations for the code attestation protocol for a single
code attestation phase we recommend to stop the protocol in case two times the received 
response $x$ (each time with a  different $nonce$) has been presented. 

\subsection{Replay Attacks}
As long as the challenge {\em nonce} is always fresh replay attacks are not possible.
Consequently the size of the nonce is a function over the lifetime of the (frequently) battery-driven prover and
the frequency of applying the code attestation protocol. For example 
if the approximate lifetime of the prover is 3 month and the verifier starts the code attestation 
protocol once per hour we state $|nonce|$ should not be smaller than four bytes (this is required to correspond to the $n$ chosen in 7.3).
The attacker can eavesdrop over the wireless all transmitted pairs $(nonce_i,x_i)$ with 
the objective to 
resend yet eavesdropped responses $x_i$. Since the nonce is the only data chunk
 providing freshness for the response computation, once a nonce $nonce_r$ is repeatedly
 transmitted by the verifier the attacker can use the time slot $\epsilon$ to upload $\widetilde{CI}$. 
However, a more realistic attack arises in case of a poor implementation of the
 'random' choice of a nonce at the verifier side.
If the attacker can infer from pairs $(nonce_1,x_1),...,$ $(nonce_r,x_r)$ to $nonce_{r+1}$ 
he can precompute $x_{r+1}$ and the time to upload and run $\widetilde{CI}$ extends from $\epsilon$ to the duration until the next run of the
code attestation protocol. However, running bogus code during the time interval of two consecutive sent challenge 
nonces $nonce_i$ and $nonce_{i+1}$ is always possible even without performing such a replay attack. 
Thus, a proper implementation of the freshness function has to ensure that the attacker cannot even infer a 
sequence of consecutive nonces $nonce_i,...,nonce_{i+j}$ with $j>1$ allowing to run bogus code undetected 
during an interval $[i,i+j]$.

\subsection{Node Depletion Attacks}
If the attacker aims at wasting the energy of the non-tamper resistant and restricted prover device 
he could masquerade as the master device and continiously send challenges $nonce$. 
Two countermeasures are possible here: firstly, one could introduce a master key $k$ which is shared
between the master device and the prover device such that $x=h_k(nonce||C_h(CI)||LAT_h||PRW)$.
The $h_k()$ denotes a keyed MAC.
However, obviously this approach contradicts with the fact that initially the attacker has 
full control over the non-tamper resistant device such that the $k$ can be read out for subsequent
depletion attacks. Due to this reason we propose a lightweight approach in which the
prover device computes and sends at maximum $n$ times per epoch a response $x$.
Here $n$ corresponds to the number of iterations recommended under 7.3.

\subsection{Other Attacks: DoS}
The protocol is not resistant against DoS attacks. To sufficiently handle depletion attacks or
attacks on the usage of the external memory an attacker can always enforce the code attestation protocol to stop.
In such situations the master device considers the code image running on the prover device as bogus.

\section{Conclusions and Open Issues}
The work at hand presents a code attestation protocol which in particular detects compression attacks
aiming to run bogus code in an undetected manner.
The code image is loaded
in a compressed manner. Only $LAT$ and $PRW$ are loaded uncompressed.
The presented approach is work in progress. Surely, more elaborated analysis are required on 
a proper choice of parameters like  $s_h$, $T_{pm}$ and $n$. Also the role of the cache needs 
to be evaluated more in depth with respect to potential security weaknesses.

\section{Acknowledgments}
The authors are most grateful to Aurelien Francillon and Claude Castelluccia who
gave insightful comments on their related work.
The work presented in this paper was supported in part by the European Commission within the STREP WSAN4CIP
of the EU Framework Programme 7 for Research and Development (http://www.ist-ubisecsens.org) as well as the German BMB+F SKIMS project. The views and conclusions contained herein are those of the
authors and should not be interpreted as necessarily representing the official policies or endorsements, either expressed
or implied, of the WSAN4CIP project, the SKIMS project or the European
Commission.


\begin{thebibliography}{12}

\bibitem{WoCh}
\newblock Wolfe, A., Chanin A. 
\newblock Executing compressed programs on an embedded RISC architecture. ACM Sigmicro Newsletter, volume 23, pp. 81-91, (1992)

\bibitem{LeBi}
\newblock Lefurgy, C., Bird, P., Chen, I., Mudge T., 
\newblock Improving Code Density Using Compression Techniques, Proceedings of the 30th annual ACM/IEEE international symposium on Microarchitecture, pp. 194-203, (1997).

\bibitem{Huf}
\newblock Huffman, D.A. 
\newblock A method for the construction of minimum redundancy codes. Proceedings of the IRE 40 (1962).

\bibitem{SWATT}
\newblock  Seshadri, A., Perrig, A., van Doorn, L., and Khosla, P. K. 
\newblock  SWATT: SoftWare-based ATTestation for embedded devices. In IEEE Symposium on Security
and Privacy (2004), IEEE Computer Society.

\bibitem{SCUBA}
\newblock  Seshadri, A., Luk, M., Perrig, A., van Doorn, L., and Khosla, P. 
\newblock  SCUBA: Secure code update by attestation in sensor networks. In WiSe '06:
Proceedings of the 5th ACM workshop on Wireless security (2006), ACM.

\bibitem{ESAS}
\newblock  Shaneck, M., Mahadevan, K., Kher, V., and Kim,Y. 
\newblock Remote software-based attestation for wireless sensors. In ESAS (2005).

\bibitem{CCS09}
\newblock Claude Castelluccia, Aurélien Francillon, Daniele Perito and Claudio Soriente,
\newblock  On the Difficulty of Software-Based Attestation of Embedded Devices, ACM CCS 2009.

\bibitem{patent}
\newblock H. Yamada, D. Fuji, Y. Nakatsuka, T. Hotta, K. Shimamura, T. Inuduka, T. Yamazaki, 
\newblock Micro-Controller for reading out compressed instruction code and program memory for compressing
instruction code and storing therein, US 6,986,029 B2

\bibitem{randomfill}
\newblock AbuHmed, T. and Nyamaa, N. and DaeHun Nyang,
\newblock Software-Based Remote Code Attestation in Wireless Sensor Network,
\newblock Global Telecommunications Conference, 2009. GLOBECOM 2009. IEEE

\bibitem{controlflowintegrity1}
\newblock Abadi, M., Budiu, M., Erlingsson, U., and Ligatti J.,
\newblock Control-flow integrity, 
\newblock In CCS'05: Proceedings of the 12th ACM conference on Computer and Communications Security (2005), ACM.

\bibitem{controlflowintegrity2}
\newblock Ferguson, C., Gu, Q., and Shi, H., 
\newblock Self-healing control flow protection in sensor applications, 
\newblock In WiSec'09 (2009), ACM.

\bibitem{DRSA}
\newblock Yang, Yi and Wang, Xinran and Zhu, Sencun and Cao, Guohong,
\newblock Distributed Software-based Attestation for Node Compromise Detection in Sensor Networks,
\newblock Proceedings of the 26th IEEE International Symposium on Reliable Distributed Systems

\end{thebibliography}
\end{document}